\documentclass{aastex}

\lefthead{Morbidelli \& Levison}
\righthead{Origin of 2000~CR$_{105}$}

\begin{document}

\title{SCENARIOS FOR THE ORIGIN OF THE ORBITS\\ 
OF THE TRANS-NEPTUNIAN OBJECTS 2000 CR$_{105}$ AND 2003 VB$_{12}$}

\author{Alessandro Morbidelli}
\affil{Observatoire de la C\^ote d'Azur\\
  Boulevard de l'Observatoire\\
  B.P. 4229, 06304 Nice Cedex 4, France}
\authoremail{morby@obs-nice.fr}

\and

\author{Harold F. Levison}
\affil{Southwest Research Institute\\ 
1050 Walnut St, Suite 426\\ 
Boulder, CO 80302 \\ \hbox{}
Submitted to {\it Astronomical Journal}}

\newpage

\begin{abstract}
  Explaining the origin of the orbit of 2000~CR$_{105}$ ($a\sim
  230$~AU, $q\sim 45$~AU) is a major test for our understanding of the
  primordial evolution of the outer Solar System.  Gladman et
  al$.$~(2001) showed that this objects could not have been a normal
  member of the scattered disk that had its perihelion distance
  increased by chaotic diffusion.  In this paper we explore four
  seemingly promising mechanisms for explaining the origin of the
  orbit of this peculiar object: (i) the passage of Neptune through a
  high-eccentricity phase, (ii) the past existence of massive
  planetary embryos in the Kuiper belt or the scattered disk, (iii)
  the presence of a massive trans-Neptunian disk at early epochs which
  exerted tides on scattered disk objects, and (iv) encounters with
  other stars.  Of all these mechanisms, the only one giving
  satisfactory results is the passage of a star. Indeed, our
  simulations show that the passage of a solar mass star at about
  800~AU only perturbs objects with semi-major axes larger than
  roughly 200~AU to large perihelion distances.  This is in good
  agreement with the fact that 2000~CR$_{105}$ has a semi-major axis
  of 230~AU and no other bodies with similar perihelion distances but
  smaller semi-major axes have yet been discovered.  The discovery of
  2003 VB$_{12}$, ($a=450$~AU, $q=75$~AU) announced a few days before
  the submission of this paper, strengthen our conclusions.
\end{abstract}


\newpage
\section{Introduction}
\label{intro}

The trans-Neptunian population of small bodies is usually divided in
two categories, the Kuiper belt and the scattered disk, although the
partition between the two is not precisely defined.  In Morbidelli et
al. (2003) we have introduced a partitioning based on the dynamics of
orbits in the current Solar System. We called {\it scattered disk} the
region of the orbital space that can be visited by bodies that have
encountered Neptune within a Hill's radius at least once during the
age of the Solar System, assuming no substantial modification of the
planetary orbits.  We then called {\it Kuiper belt} the complement of
the scattered disk in the $a>30$~AU region.

The bodies that belong to the scattered disk in this classification
scheme do not provide us with any significant clue about the
primordial architecture of the Solar System.  This is because their
current orbits can be achieved by purely dynamical evolution in the
current planetary system from objects that started in nearly-circular
nearly-coplanar orbits in Neptune's zone. The opposite is true for the
orbits of the Kuiper belt objects.  All bodies in the Solar System
must have been formed on orbits typical of an accretion disk (e.g.
with very small eccentricities and inclinations).  Therefore, the fact
that most Kuiper belt objects have a non-negligible eccentricity
and/or inclination reveals that some excitation mechanism, which is no
longer at work, was active in the past.

In this respect, particularly interesting are the Kuiper belt bodies
with large semi-major axis ($a>50$~AU), such as 2001~QW$_{297}$
($a=51.3$~AU, $q=39.5$AU, $i=17.1^\circ$), 2000~YW$_{134}$
($a=58.4$~AU, $q=41.2$AU, $i=19.8^\circ$), 1995~TL$_8$ ($a=52.5$~AU,
$q=40.2$AU, $i=0.2^\circ$), 2000~CR$_{105}$ ($a=230$~AU, $q=44.4$~AU,
$i=22.7^\circ$) and, last discovered, 2003~VB$_{12}$ ($a=531$~AU,
$q=74.4$~AU, $i=11.9^\circ$; Brown et al$.$~2004).  We call these
objects {\it extended scattered disk} objects for three reasons:
{\it (i)} they do not seem to belong (some caution is needed because
of the uncertainties in the orbital elements of these objects) to the
scattered disk according to our definition but are very close to its
boundary (Gladman et al., 2001; Emel'yanenko et al., 2003; Morbidelli
et al., 2004); {\it (ii)} some of these bodies have sizes of several
hundred kilometers, suggesting that they formed much closer to the
Sun, where the accretion timescale was sufficiently short (Stern,
1996) and were subsequently transported to these current locations;
{\it (iii)} the lack of objects with $q>41$~AU and $50<a<200$~AU
cannot be due to observational biases (given that many classical belt
objects have been discovered up to distances of 45--50~AU), suggesting
that these extended scattered disk objects are {\it not} the
high-eccentricity members of an excited belt beyond 50~AU.  These
considerations indicate that in the past the scattered disk extended
beyond its present boundary in perihelion distance.

Perhaps that most promising idea for the formation of these extended
scattered disk objects (or at least most of them) was recently studied
by Gomes (2003b). Gomes (2003b) investigated whether the scenario
proposed in Gomes (2003a) for the origin of the `hot Kuiper belt
population' (the population of non-resonant bodies with large
inclinations) could also be responsible for the extended scattered
disk. We remind the reader that in Gomes's scenario, the hot
population was originally a part of the primordial, massive scattered
disk population.  During Neptune's migration, a small fraction of
these objects had their perihelion distances increased and thus they
became permanently trapped on stable orbits. Gomes (2003b) found
several particles that increased perihelion distance well beyond
$40$~AU in the $a>50$~AU region.  However, in all cases, the
semi-major axis was smaller than 200~AU.  Therefore, Gomes's mechanism
implies the existence of bodies with $q\!sim\!45\,$AU spread from
$a\!\sim\!50\,$AU to $\sim\!200\,$AU, but no objects has been
discovered at the small semi-major axis end of this range.  This, in
spite of the fact that observational biases favor the discovery of
small semi-major axis objects.

Indeed, 2000~CR$_{105}$ is special for a couple of reasons.  Until the
recent discovery of 2003~VB$_{12}$, it had the largest semi-major axis of
the extended scattered disk, by a large margin.  It also had a
significantly larger perihelion distance than any other extended
scattered disk object.  Although it is possible that 2000~CR$_{105}$
is just an outlaying member of the extended scattered disk, the fact
that no objects with perihelion distance comparable to that of
2000~CR$_{105}$ but with a smaller $a$ had been discovered seemed
significant to us.  This is particularly true considering that
observational biases sharply favors the discovery of objects with
smaller semi-major axes.  Thus, we were motivated to look for
dynamical mechanisms that preferentially raised the perihelion
distance of scattered disk objects at large semi-major axis. The
discovery of 2003~VB$_{12}$ came a few days before the submission of this
paper, and confirmed that our investigation was well motivated.  In
fact, the orbit of this body definitely falls beyond the distribution
produced in Gomes model.

Some of the mechanisms investigated in this paper have been already
suggested by Gladman et al. (2001), but never have been quantitatively
explored.  In Section~\ref{nep_ecc} we consider the case where Neptune
was more eccentric in the past, as proposed by Thommes et
al$.$~(1999).  It is obvious that a more eccentric Neptune would
produce a extended scattered disk, but it is not known, {\it a
  priori}, what eccentricity would be required to produce objects on
2000~CR$_{105}$-like orbits, and over what timescale.  In
Section~\ref{planet} we investigate the effects of the presence of
terrestrial mass planet(s) in the Kuiper belt or in the scattered
disk, as proposed by Morbidelli and Valsecchi~(1997) and Brunini and
Melita~(2002).  In Section~\ref{dtides} we propose a new model for the
origin of 2000~CR$_{105}$, in which the tides raised by a massive disk
beyond $\sim 70$~AU increased the perihelion distance of high inclined
scattered disk objects.  Finally, in Section~\ref{senc} we investigate
the stellar passage scenario.  This scenario has been first proposed
by Ida et al. (2000) to explain the structure of the inner Kuiper
belt.  Although we disagree that the all of the sculpting of the
Kuiper belt could be due this mechanism (see Levison et al$.$, 2004),
it is still possible that a more gentle encounter could have formed
objects like 2000~CR$_{105}$.

\section{Eccentric Neptune}  
\label{nep_ecc}
  
It is possible that at sometime in the early epochs of the Solar
System, Neptune was on an orbit that was significantly more eccentric
than its current one.  A high eccentricity could have been achieved
during a phase when the planet experienced encounters with Jupiter and
Saturn, as proposed by Thommes et al. (1999). It could also be the
result of interactions between Neptune and other hypothetical massive
planetary embryos or of its temporary capture in a resonance with one
of the other planets, although these scenarios have never been
quantitatively simulated.  In this section we investigate the effects
that an eccentric Neptune would have on the formation of the scattered
disk.

Our numerical experiment is very simple. We have performed a series of
$1\,$Gyr integrations of the evolution of 1000 test particles,
initially placed on circular and coplanar orbits between 30 and 50~AU.
The runs differ from one another in the eccentricity of Neptune, which
was set to either 0.1, 0.2, 0.3 and 0.4. All the other initial orbital
elements of the planets have been chosen to be equal to their current
values.  However, Uranus was removed in the integrations where
Neptune's eccentricity was equal to 0.3 or 0.4, to avoid close
encounters between the planets.  The integrations have been done using
the swift$\_$rmvs3 integrator (Levison and Duncan, 1994) with a global
timestep of 1 year.  

\begin{figure}
\epsscale{0.4}
\plotone{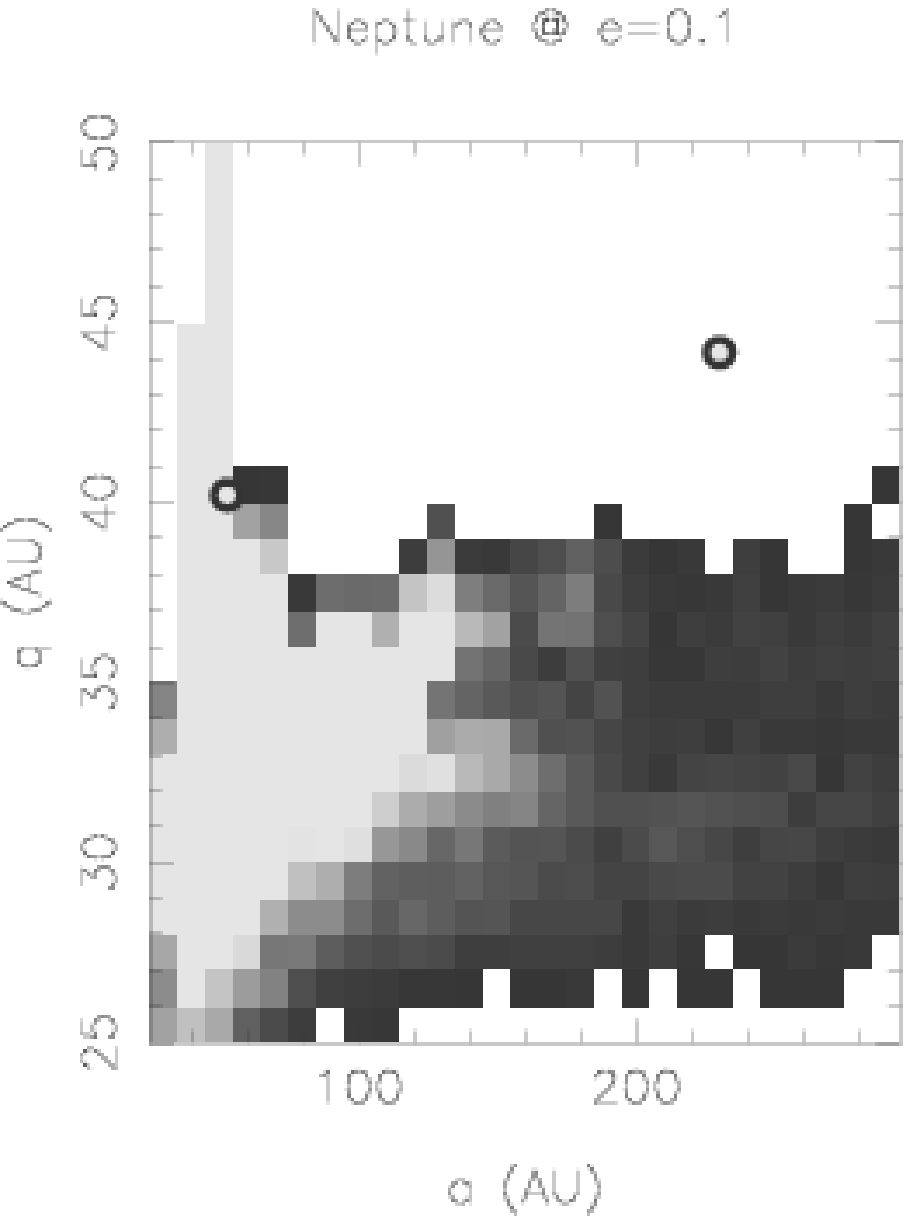}
\plotone{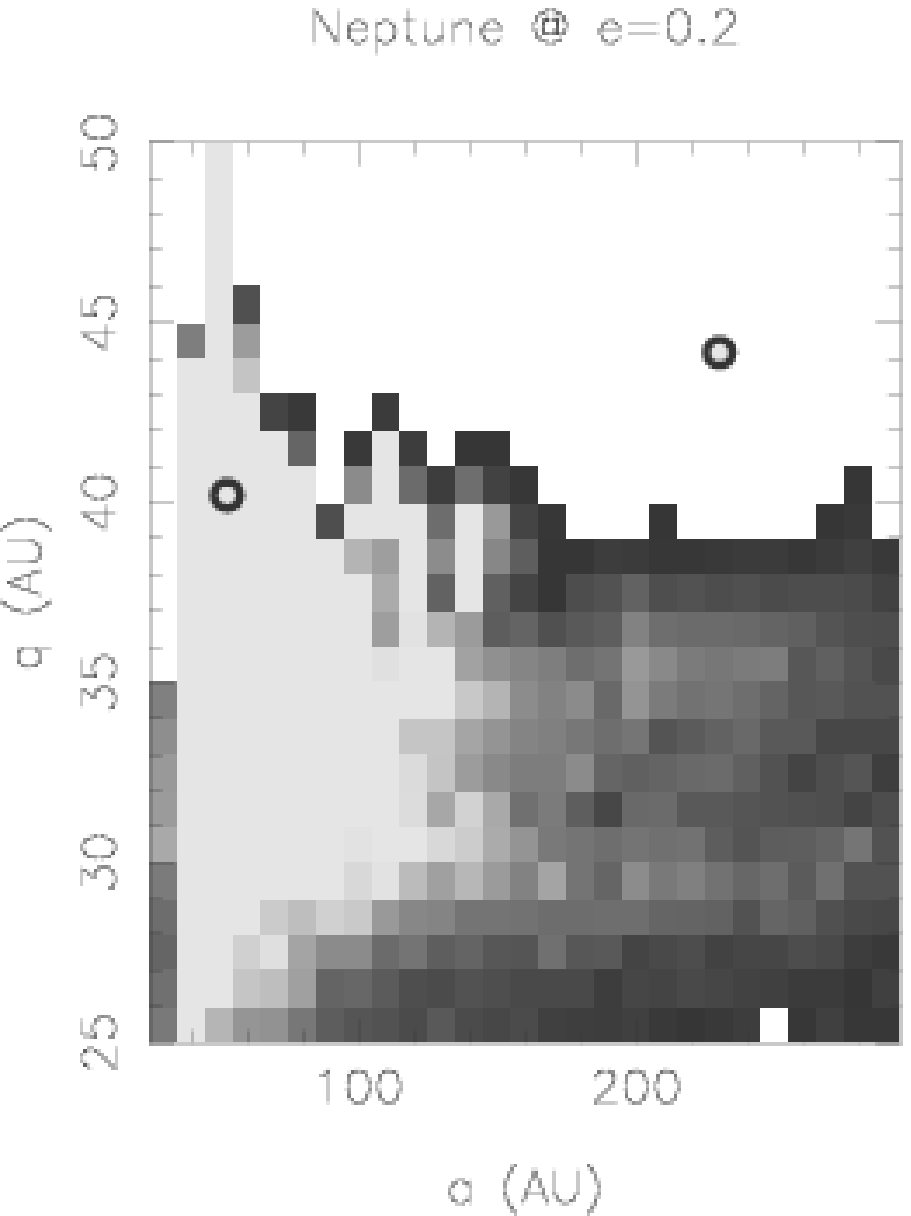}
\plotone{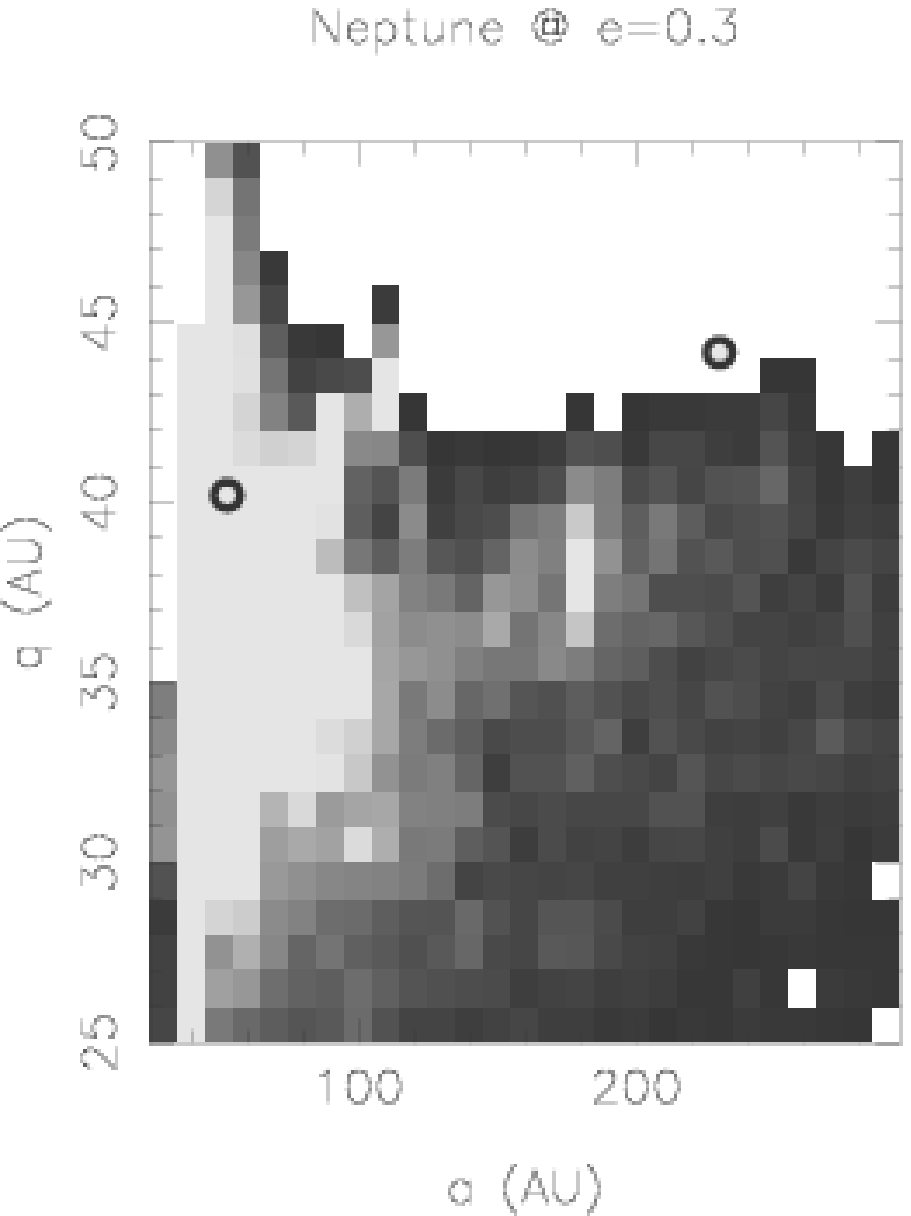}
\plotone{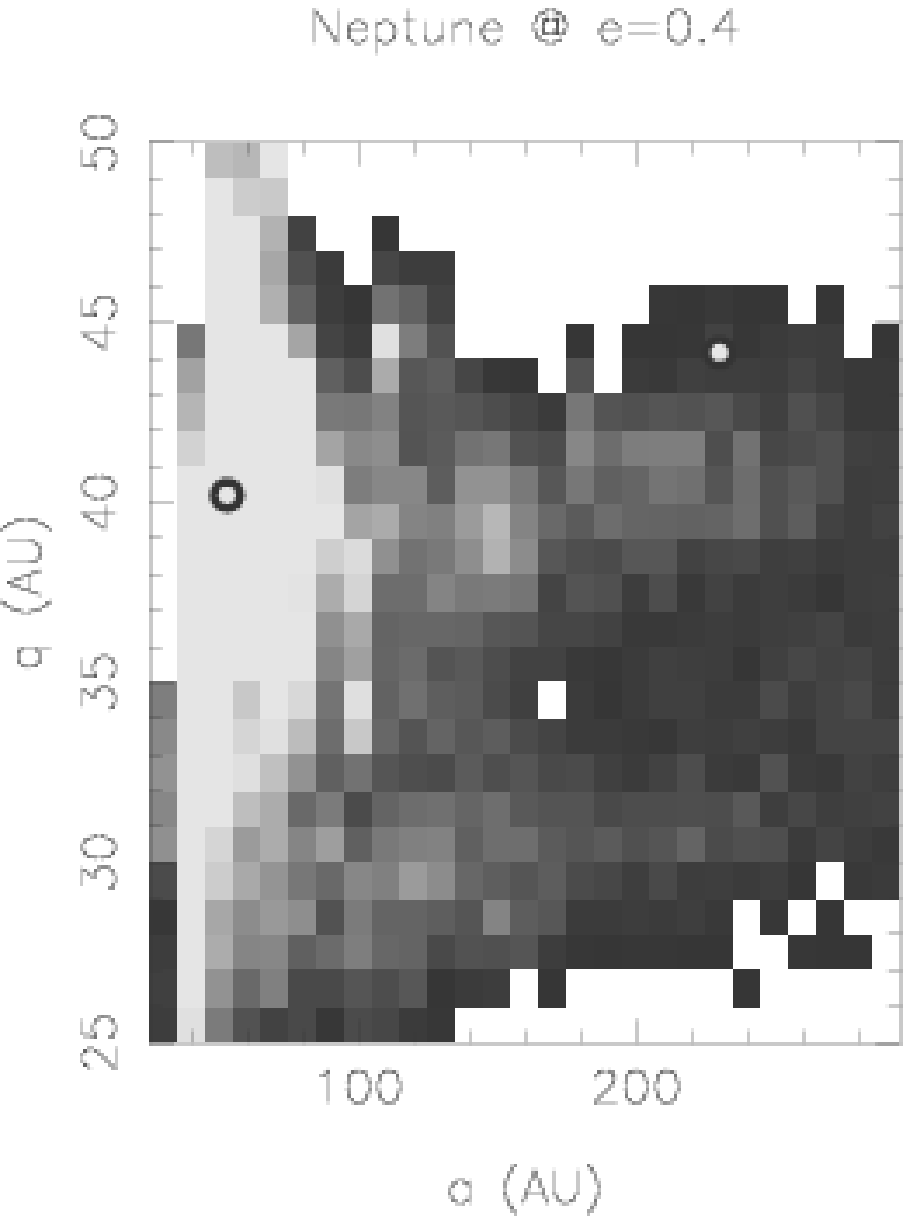}
\caption{The extent of the scattered disk that is generated by a Neptune
  on increasingly eccentric orbits. The gray scale denotes the
  cumulative time spent by the integrated test particles in the
  $10\times 2$~AU cells of the $a,q$ plane. A darker color denotes a
  shorter time.  The uncolored region is the one that has never been
  visited by a test particle during the 1~Gy integration. The gray
  open dots, bounded by black rings, denote the position of
  1995~TL$_8$ ($a=52.69$~AU, $q=40.2$~AU) and 2000~CR$_{105}$
  ($a=230$~AU, $q=44.2$~AU).}
\label{Nept-ecc} 
\end{figure}

To visualize the extent of the scattered disk produced in the above
simulations, we have divided the $a,q$ plane in cells and computed the
cumulative time spent by each test particle in each cell.  The results
are illustrated in the 4 panels of Fig~\ref{Nept-ecc}, using a gray
scale where a darker color corresponds to a shorter residence time.
The white areas shows the regions which were not visited by any test
particle during the entire integration time. The gray dots surrounded
by black rings denote the current positions of 1995~TL$_8$ and
2000~CR$_{105}$, prominent representatives of the extended scattered
disk. As one sees, while objects on orbits similar to 1995~TL$_8$ are
easily produced by a Neptune on an orbit with $e=0.1$ (but not if
Neptune were on its current orbit; Duncan and Levison, 1997), objects
on orbits like 2000~CR$_{105}$ require that Neptune's eccentricity is
at least $0.4$.  Objects with orbits similar to 2003~VB$_{12}$ are not
produced even in this extreme case.

Although it is possible that Neptune once had an eccentricity as large
as $0.4$ (see Thommes et al., 1999), we doubt that this scenario can
explain the origin of 2000~CR$_{105}$'s orbit for two reasons. First,
this scenario predicts many more bodies with $50<a<90$~AU than with
$200<a<240$~AU, for $40<q<45$~AU.  This can be seen in the bottom
right panel of Fig.~\ref{Nept-ecc}, which shows the total time spent
in the former region is much larger than in the latter.  This result
is exacerbate by the observational biases which would strongly favor
the discovery of the bodies with the smallest semi-major axes.

The second, even more compelling reason, is that in our simple
integrations it takes 92~My before that the first body reaches
$a>220$~AU and $q>44$~AU (it takes 24~My to reach $a>220$~AU, without
restriction on $q$).  However in reality, it is not possible for
Neptune's eccentricity to have remained this large for so long.  In a
more realistic situation, Neptune's eccentricity is damped very
rapidly (less than a million years) by the dynamical friction exerted
by the planetesimal disk (Thommes et al., 1999).  In fact, in none of
the Thommes et al$.$ integrations a body on a 2000~CR$_{105}$-like
orbit was ever produced (Thommes, private communication).

Therefore, we conclude that an eccentric Neptune at early epochs
cannot be a plausible explanation of the origin of 2000~CR$_{105}$ or
2003~VB$_{12}$.

\section{Rogue planet}
\label{planet}

Morbidelli and Valsecchi (1997) and Petit et al. (1999) have proposed
that an Earth--mass body, scattered outward by Neptune, might have
caused the orbital excitation observed in the trans-Neptunian region.
More recently Brunini and Melita (2002) proposed that a planet on a
moderate eccentricity orbit with $a\sim 60$~AU could explain the
putative edge of the Kuiper belt at $\sim 50$~AU (Allen et al$.$~2001;
Trujillo and Brown, 2001).  Although detailed investigations seem to
indicate that these scenarios (often nicknamed the ``rogue planet
scenarios'') cannot be responsible for the observed Kuiper belt
structure (see Morbidelli et al., 2003 for a discussion), it is worth
briefly investigating whether a rogue planet in the Kuiper belt or in
the scattered disk could explain the origin of 2000~CR$_{105}$ and/or
2003~VB$_{12}$.  

We consider a scenario similar to that proposed by Petit et
al$.$~(1999), who speculated on the existence of massive bodies in the
scattered disk during the early epochs of the Solar System.  To
accomplish this, we followed the evolution of a system containing the
4 giant planets, a number of embryos initially between Uranus and
Neptune, and 983 test particles for 1 billion years.  In order to put
ourselves in the most favorable position to generate objects on orbits
like 2000~CR$_{105}$ and 2003~VB$_{12}$, we have considered the extreme
and unrealistic case of initially having 10 half-Earth mass embryos in
the system.  The test particles were initially placed between 25 and
35~AU or between 40 and 50~AU, on quasi-circular co-planar orbits.

During the simulation, all of the embryos at some point found
themselves in the scattered disk, with $a>30$~AU. Of them, six
temporarily reached a semi-major axis larger than 100~AU.  Some of the
embryos remained in the system for a long time, where they could
presumably scatter the test particles.  Indeed, five embryos had a
lifetimes longer than 100~My, and two survived to the end of the
integration.  The surviving embryos were still on a Neptune-crossing,
however, so they are presumably not stable.

\begin{figure}
\epsscale{1.0}
\plottwo{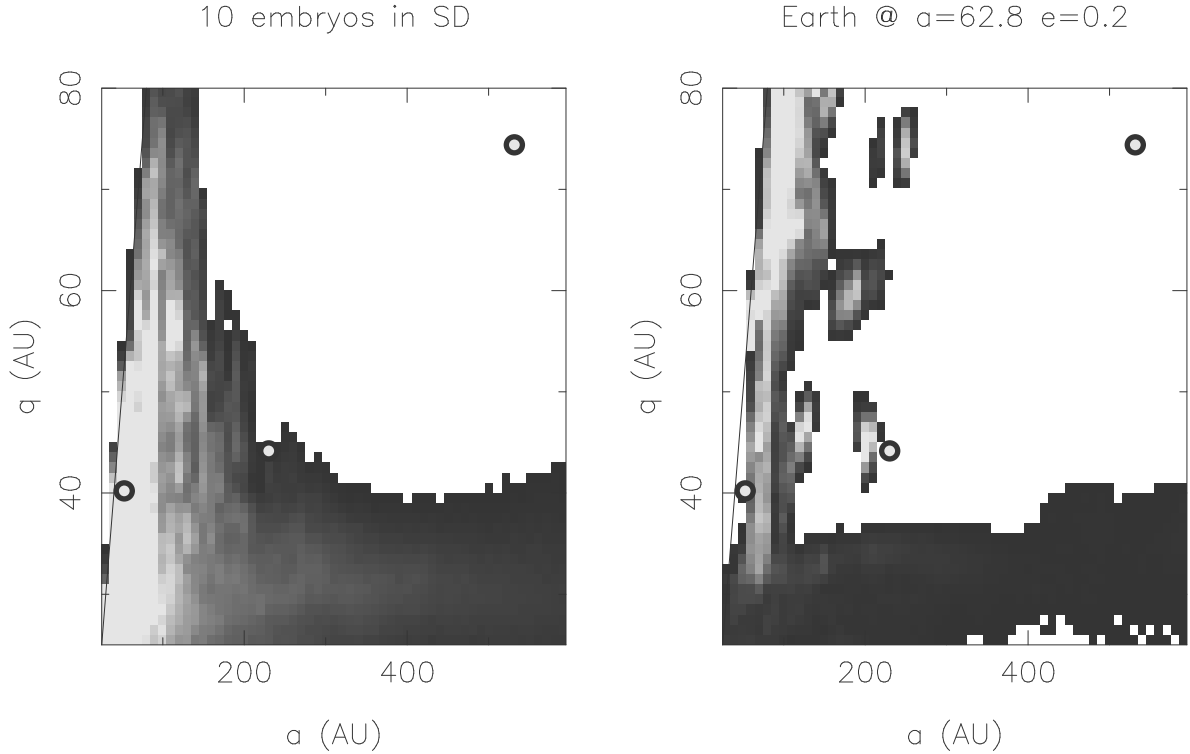}{dist-pl.ps}
\caption{Left: the region of the $a,q$ plane visited by test particles
  evolving under the influence of the 4 giant planets and 10
  half-Earth mass embryos.  The embryos were initially between Uranus
  and Neptune and all eventually evolved into the scattered disk.  The
  gray scale representation is analog to that of Fig.~\ref{Nept-ecc}.
  Right: the same, but for particles initially between 42 and 75 AU,
  under the influence of the 4 giant planets and of an Earth--mass
  planet at $a=62.83$~AU, $e=0.2$ and $i=6^\circ$. The dot in the
  upper right coner of each panel represents 2003~VB$_{12}$.}
\label{rogue} 
\end{figure}

The left panel of Fig.~\ref{rogue} shows the $(a,q)$ region covered by
the test particles during the simulation.  It was generated using the
procedures described above for generating Fig.~\ref{Nept-ecc}.  The
region visited by our test particles marginally overlaps the orbit of
2000~CR$_{105}$.  However, if this scenario were correct, we would
expect many more objects with perihelion distances similar to
2000~CR$_{105}$, but with smaller semi-major axis.  This problem is
not alleviated by considering only particles that survive for a long
time in the simulation.  Moreover, the orbit of 2003~VB$_{12}$ is very
far from the boundary of this distribution.  Thus we conclude that
also the Petit et al$.$~(1999) model in principle cannot explain these
two objects.

We have also considered an Earth-mass planet initially on an orbit
similar to that postulated by Brunini and Melita (2002), with
$a=62.83$~AU, $e=0.2$ and $i=6^\circ$.  This planet has the advantage
of having an aphelion distance at $\sim 75$~AU, which is very close to
the perihelion distance of 2003~VB$_{12}$.  It therefore seems to be a
good candidate to emplace objects at the locations of both
2000~CR$_{105}$ (which would be deeply planet-crosser) and
2003~VB$_{12}$.

We have integrated for 4~Gy the orbits of 100 test particles initially
on circular and coplanar orbits between 60 and 90~AU under the
gravitational influence of the Sun, the 4 giant planets, and the rouge
planet.  The result is shown in the right panel of Fig.~\ref{rogue},
with the same representation used in Fig.~\ref{Nept-ecc}.  Unlike in
the previous plots of this paper, only the last 2~Gy of evolution are
used to compute of the region covered by the particles. The orbit of
2000~CR$_{105}$ is reproduced!  Moreover, for the particles with
$40<q<50$~AU the semi-major axis distribution peaks nicely at 200~AU.
Thus this mechanism is consistent with the fact that we have not found
objects with the same $q$ as 2000~CR$_{105}$ but with smaller
semi-major axes.

However, we caution that our test particle density distribution near
the position of 2000~CR$_{105}$ is due to a {\it single} particle,
which is scattered into that region at $t\!=\!720\,$Myr and then
evolves in a quasi-stable manner for the age of the Solar System.
Hence our apparently nice result above suffers from small number
statistics.

However, the orbit of 2003~VB$_{12}$ remains well beyond the reach of the
particles scattered by the rogue planet. Even on a timescale of 4~Gy,
an Earth-mass planet has difficulty transporting objects much further
than $a\!\sim\!250$ AU.  The bodies that reach $a>250$~AU have
$q<40$~AU.  Their small perihelion distance indicates that Neptune,
rather than the rogue planet, is responsible for their large
semi-major axes.  In fact, the only way an object reached large
semi-major axes in our integrations was to first have its perihelion
distance driven down to Neptune's orbit due to interactions with the
rogue planet, and then Neptune drove it to large $a$.

This simulation shows that the naive expectation that a planet would
populate the entire orbital region that crosses its own orbit is not
correct.  An Earth-mass planet at the edge of the Kuiper belt simply
cannot transport objects from a nearly circular orbit to large
semi-major axes in the age of the Solar system without first handing
them off to Neptune.

Having said this, given the small number of experiments we have thus
far performed, we, of course, cannot rule out that there is a
combination of planet mass and distance that can produce both
2000~CR$_{105}$ and 2003~VB$_{12}$.  There is a huge parameter space of
possibilities that cannot be exhaustively covered in a practical way.
However, the results presented in this section indicate that the most
intuitive and often invoked solutions do not always trivially work.

In addition, our results show that it would take a long time to create
an object like 2003~VB$_{12}$.  So any trans-Neptunian planet that could
make this object would, most likely, need to be in the Solar System
today.  The presence of one or more planets in the distant Solar
System would raise sever questions such as: how did these planets form
so far from the Sun?  How were these planets transported to their
current distant location?  Is their formation or transport compatible
with the observed properties of the Kuiper belt and with the orbital
distribution of the other giant planets?  Why haven't these planets
yet been observed?  Science should always give preference to the most
simple theories --- ones that do not raise more problems than they
solve. We are convinced that the rogue planet scenario does not fall
into this category.  A much more credible scenario for the origin of
2000~CR$_{105}$ and 2003~VB$_{12}$ is presented in \S{\ref{senc}}.

\section{Disk tides}
\label{dtides}

In this section we present a wholly new idea for the formation of
2000~CR$_{105}$ --- one which, unfortunately, fails to work.  We
present this mechanism for completion and because we believe that the
dynamics presented here could be of use in the future.

Imagine that a massive and dynamically cold trans-Neptunian disk of
planetesimals persisted for a long time. This disk would have exerted
tidal forces on bodies with large semi-major axes and moderate to
large inclinations, similar to those exerted by the Galactic disk on
Oort cloud comets.  As a consequence, as scattered disk bodies evolved
outward, they would have entered a phase where their inclinations and
perihelion distances would have been oscillating due to the presence
of this disk.  If this disk, or at least part of it, dispersed while
some objects were in this phase, some of them would have been left
with large perihelion distances.

\begin{figure}
\epsscale{1.0}
\plottwo{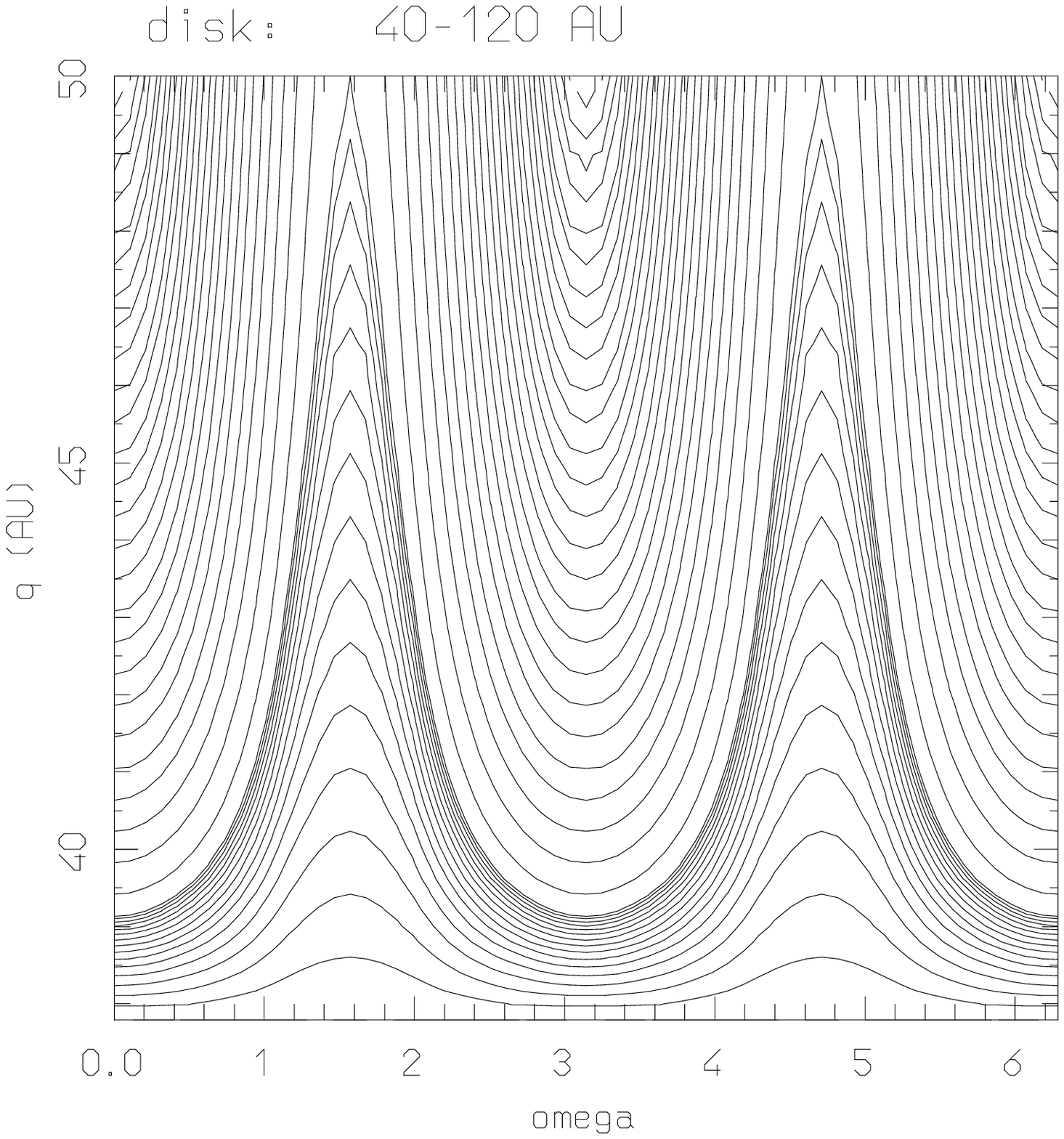}{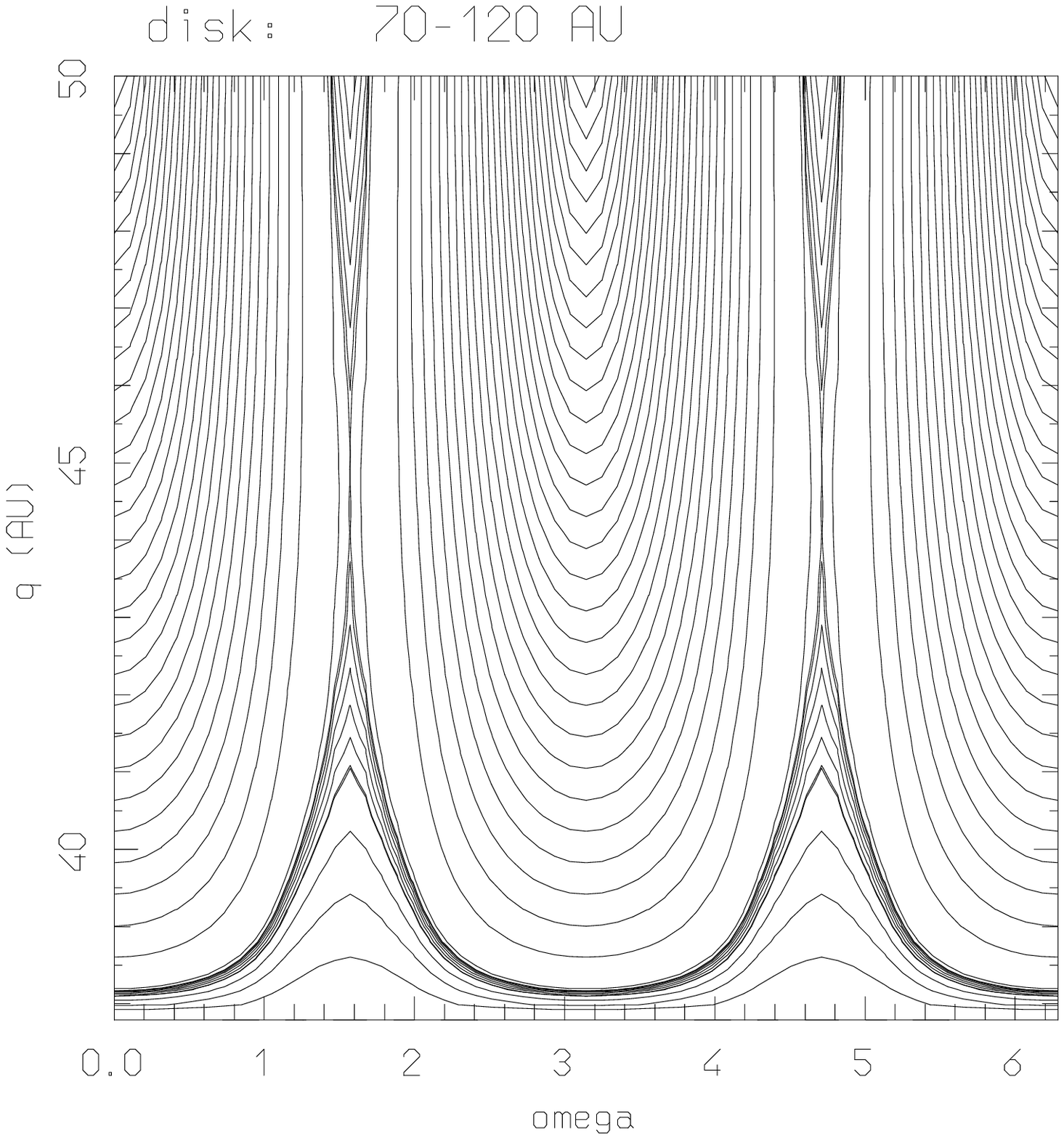}
\caption{The secular $\omega,q$ evolution induced by the 4 giant planets and
  a disk of: (left) $94 M_\oplus$ between 40 and 120~AU, (right): $46
  M_\oplus$ between 70 and 120 AU. Both panels are computed for small bodies
  with $a=230$~AU and $H=\sqrt{a(1-e^2)}\cos i=8.329$ (the current value of
  2000~CR$_{105}$).}
\label{kozai} 
\end{figure}

The secular evolution induced on small bodies by the 4 giant planets
(assumed to be on coplanar and circular orbits) and a massive disk
situated on the planets' orbital plane can be analytically computed
with a trivial adaptation of the approach usually followed to compute
the effects of the Kozai resonance (see Thomas and Morbidelli, 1996;
chapter~8 of Morbidelli, 2002). The motions of the eccentricity and of
the argument of perihelion $\omega$ are coupled, while the semi-major
axis and the quantity $H=\sqrt{a(1-e^2)}\cos i$ remain constant.
Fig.~\ref{kozai} shows the possible trajectories on the $\omega,q$
plane for $a=230$~AU and $H=8.329$, which are the values corresponding
the current orbit of 2000~CR$_{105}$, once its inclination is computed
with respect invariant plane of the 4 giant planets. The massive disk
is assumed to extend from 40 to 120 AU in the case illustrated by the
left panel and from 70 to 120~AU in the case illustrated by the right
panel. Both disks had the same surface density, $\Sigma \propto
r^{-2}$, which is a simple extrapolation of the surface density of
solid material in the giant planets region (Weissman and Levison,
1997).  Thus, the total mass of the disk on the left was $94
M_\oplus$, while the total mass of the disk on the was $46 M_\oplus$.

As one sees, as the inner edge of the disk moves outwards from 40 to
70 AU, the libration region increases in width. Consider now an
initial conditions with $q\sim 38$~AU at $\omega=0$. If the inner edge
of the disk is at 40~AU, these initial conditions give orbits that
have only moderate oscillations of perihelion distance while $\omega$
precesses. If the inner edge of the disk is at 70~AU, they give orbits
which are in the libration region and along which $q$ eventually
increases beyond 44~AU. The timescale for this increase is of order of
a few million years.  For bodies with higher inclination (smaller
value of $H$) than 2000~CR$_{105}$, the change in perihelion distance
is enhanced, while for bodies with smaller inclination it is less
pronounced.

From these results, the scenario that we tentatively envision is the
following.  Neptune dispersed the bodies in its vicinity forming a
scattered disk; assuming that the planet was more or less on its
current orbit, the perihelion distances of the scattered disk bodies
with large semi-major axis ranged up to $\sim 38$~AU. Because of the
tides generated by the massive disk, the scattered disk bodies with
moderate or high inclinations suffered perihelion distance
oscillations, coupled with the precession of their perihelion argument
$\omega$.  Meanwhile, the collisional and dynamical erosion of the
massive disk caused an effective outward migration of the disk's inner
edge.  As a consequence, the amplitude of the perihelion distance
oscillations increased, and bodies with $a\sim 230$~AU and $H\lesssim
8.329$ (2000~CR105 values) were eventually captured in the region of
phase space where $\omega$ librates.  As a result, these objects went
through phases where their perihelion distance could get up to 44~AU
or more.  Eventually, the entire disk lost its mass, perhaps through
collisional erosion, and the tide disappeared.  Consequently $\omega$
started to circulate again the perihelion distances of the bodies
remained essentially frozen for the rest of the Solar System's
lifetime.

We have attempted to simulate this scenario with a numerical
integration.  We followed the evolution of 300 scattered disk objects
under the gravitational influence of the Sun, 4 giant planets, and a
$96\,M_\oplus$ disk spread between 40 and 150~AU.  The disk is divided
in two parts: the inner part encompassing the region between 40 and
60~AU, while the outer part encompassing the region beyond 60~AU.  The
mass of the inner part linearly decays to zero in 70~My, while that of
the outer part decays in 300~My.  The initial conditions for the test
particles are a subset of the initial conditions in Levison and Duncan
(1997).

\begin{figure}
\epsscale{1.0}
\plotone{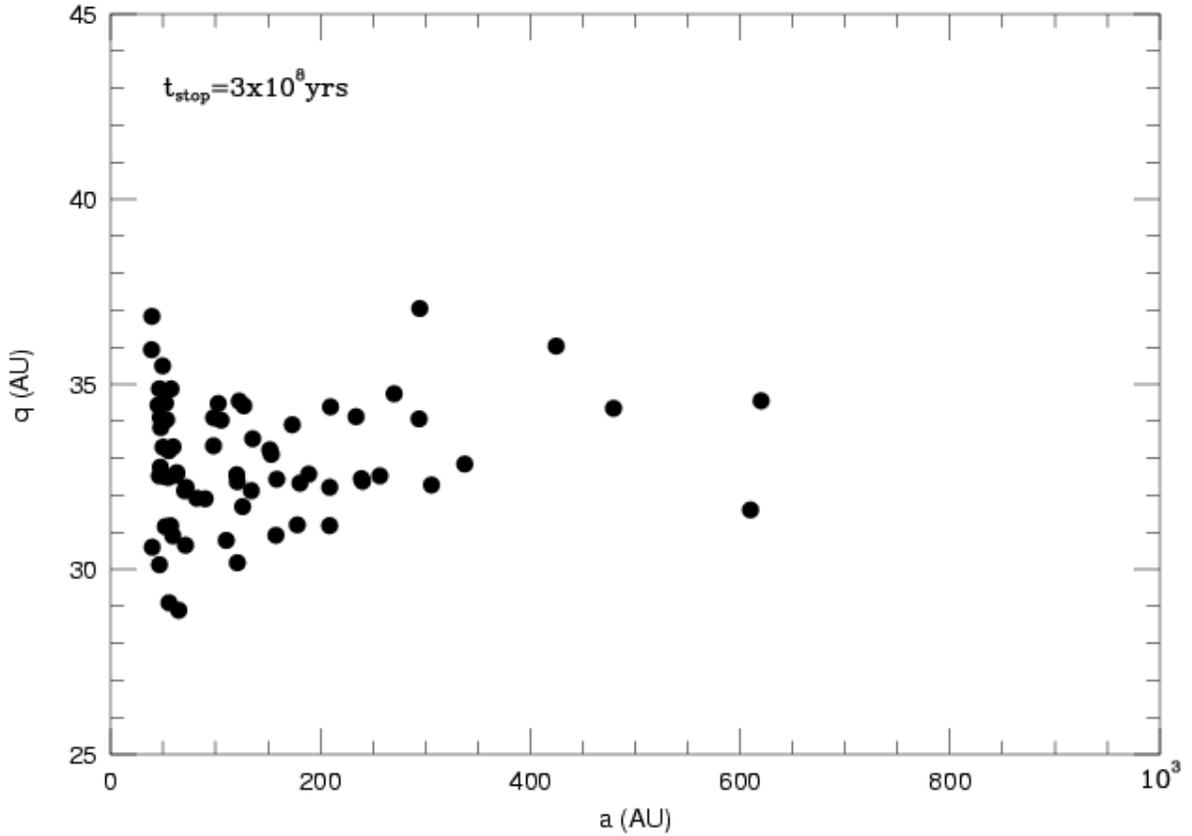}
\caption{The distribution, after 300~My, of a set of scattered
  disk particles, which evolved under the gravitational influence of
  the 4 giant planets and a massive trans-Neptunian disk. The disk has
  initially 96 Earth masses between 40 and 150 AU; its inner part (41
  Earth masses between 40 and 60~AU) is eroded in 70~My while the
  remaining outer part is eroded in 300~My. No particles are found on
  orbits typical of the extended scattered disk.}
\label{deception} 
\end{figure}

We followed the evolution of these particles with a version of the
swift$\_$rmvs3 integrator modified so that the gravitational potential
of the two parts of the disk were added to the equations of motion of
both the planets and the particles.  The orbital distribution of the
particles at the time when the disk is totally dispersed is shown in
Fig.~\ref{deception}. None of the particles have a perihelion distance
larger than 38~AU. We believe this result is due to the fact that
close encounters with Neptune are so frequent that the particles do
not have time enough to respond to the slow, secular forcing of the
disk tide. As a test, we performed a new simulation where we removed
the giant planets and kept the mass of the disk constant.  We indeed
observed that the particles with inclinations larger than 30 degrees
and semi-major axes in the 200--260~AU region had their perihelion
distances lifted above 45~AU, in good agreement with the analytic
estimates. Therefore, we are forced to conclude that the disk tide
scenario for the origin of a high inclined extended scattered disk
does not work.

\section{Stellar encounter}
\label{senc}

Observing that the dynamical excitation in the Kuiper belt apparently
increases with semi-major axis, Ida et al. (2000) suggested that this
structure might record the hyperbolic passage of a solar mass star at
100--200~AU from the Sun. With improved data, it now seems unlikely
that the complexity of the orbital structure of the Kuiper belt can be
explained by a stellar passage.  However, the truncation of the Kuiper
belt at $\sim 50$~AU might still be caused by such a passage
(Kobayashi and Ida, 2001; Melita et al., 2002; Levison et al., 2004).
The details of such an encounter are described in Levison et al$.$
(2004, hereafter LMD04).


In this section we investigate whether a stellar encounter could be
responsible for placing 2000~CR$_{105}$ and 2003~VB$_{12}$ on their
current orbits.  In particular, since these objects are so large and
thus unlikely to have formed at their current semi-major axes
(Stern~1996), we will study whether a passing star could be deliver
them from an early massive scattered disk.

We follow the procedures described in detail in LMD04, but which we
briefly review here: $1)$ We started with a simulation of the
formation of the Oort cloud by Dones et al$.$~(2004).  From this
simulation we have the total time history of the Oort cloud formation
according to this model.  $2)$ We extracted the position of planets
and particles from the DLDW04 calculations a specific time. $3)$ We
integrated the orbits of these particles during a stellar encounter.
Since this work is intended as a proof of concept, we restricted
ourselves to a `typical' $1\,M_\odot$ on a hyperbolic orbit with
$\omega=90^\circ$, $i=45^\circ$ and an unperturbed encounter velocity
of 0.2 AU/y.  The only characteristic of the encounter we vary is the
perihelion distance of the star.

\begin{figure}
\epsscale{1.0}
\plotone{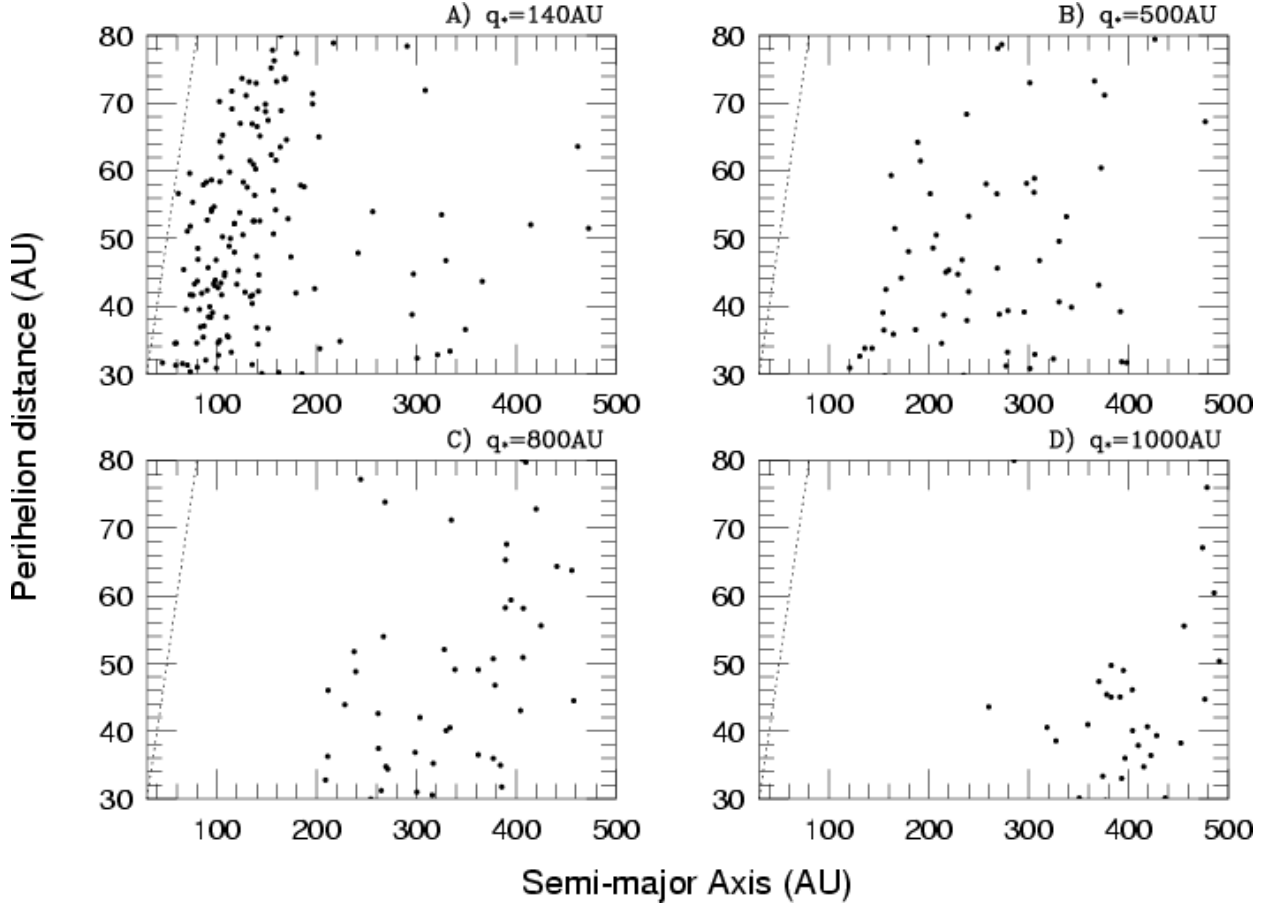}
\caption{The extended scattered disk that resulted from a series of
  passing stars.  In all cases the passing star was $1\,M_\odot$ and
  was on a hyperbolic orbit with $v_\infty\!=\!0.2\,$AU/yr,
  $\omega=90^\circ$, $i=45^\circ$.  The only thing that varies from
  panel to panel is the star's perihelion distance, $q_\star$.  The
  particles were initially in the scattered disk that was created
  during Dones et al$.$~(2004) simulations of Oort cloud formation. In
  particular, we took the scattered disk at $10^5$ years into the
  Dones et al$.$ simulation, but our results are not significantly
  effected by this choice.  See LMD04 for more detail. A)
  $q_\star\!=\!140\,$AU. B) $q_\star\!=\!500\,$AU. C)
  $q_\star\!=\!800\,$AU. D) $q_\star\!=\!1000\,$AU.}
\label{cr-stars} 
\end{figure}

Fig.~\ref{cr-stars} shows the results of our simulations for four
different values of the star's perihelion distance ($q_\star$): 140~AU,
500~AU, 800~AU, and 1000~AU.  In all cases but the
$q_\star\!=\!1000\,$AU run, objects like 2000~CR$_{105}$ and
2003~VB$_{12}$ are created.  The $q_\star\!=\!1000\,$AU run is simply too
weak to produce these objects.

However, as we explain in \S{\ref{intro}}, we believe that one of the
important characteristics that we need to explain with these models is
the dearth of observed objects with perihelion distances comparable to
2000~CR$_{105}$, but with smaller semi-major axes.  If this is indeed
the case, then we can put some constraints on $q_\star$.  Small
perihelion passages, like the ones required to sculpt the outer edge
of the Kuiper belt (LMD04), can be ruled out because they tend place
too many objects on large $q$ orbits close or interior to $100\,$AU.
The run shown in Fig.~\ref{cr-stars}A, for example, has 17 objects
with $42\!<\!q\!<\!48\,$AU and $a\!<\!150\,$AU, while only 6 in the
same range of $q$ but with $a\!>\!150\,$AU.  Since observational
biases tend to favor the discovery of objects with smaller semi-major
axes, it is difficult to reconcile this model with the observations.
If the edge of the Kuiper belt was really formed by a stellar
encounter at $\sim 150$~AU, the event that placed 2000~CR$_{105}$ and
2003~VB$_{12}$ onto their current orbits must necessarily have occurred
afterwords.

At larger $q_\star$'s the models begin to look like the distribution
that we believe that the data is indicating.  In both the
$q_\star\!=\!500\,$AU run and the $q_\star\!=\!800\,$AU run there is s
sharp transition interior to which there are no objects with large
$q$.  But, exterior to this boundary the star strongly perturbed the
scattered disk and many objects were lifted to $q\!\sim\!45\,$AU or
beyond.  This sharp transition in semi-major axis between perturbed
and non-perturbed bodies was already observed in Kobayashi and Ida
(2001). Similar results can be found in Fernandez and Brunini (2000).
For the stars studied here, the best fit to the data is
$q_\star\!\sim\!800\,$AU.

LMD04 set a tight constraint on the time when the putative stellar
encounter that truncated the Kuiper belt could have occurred, by
looking at the ratio between the scattered disk population and the
extended scattered disk population in the 50--100~AU region.
Unfortunately, we cannot repeat the same exercise here, because this
more distant encounter affected only the bodies with $a>200~AU$, and
in this region the number of {\it known} objects in both the scattered
disk and the extended scattered disk is still too limited for
statistical considerations. 

\begin{figure}
  \epsscale{1.0} \plotone{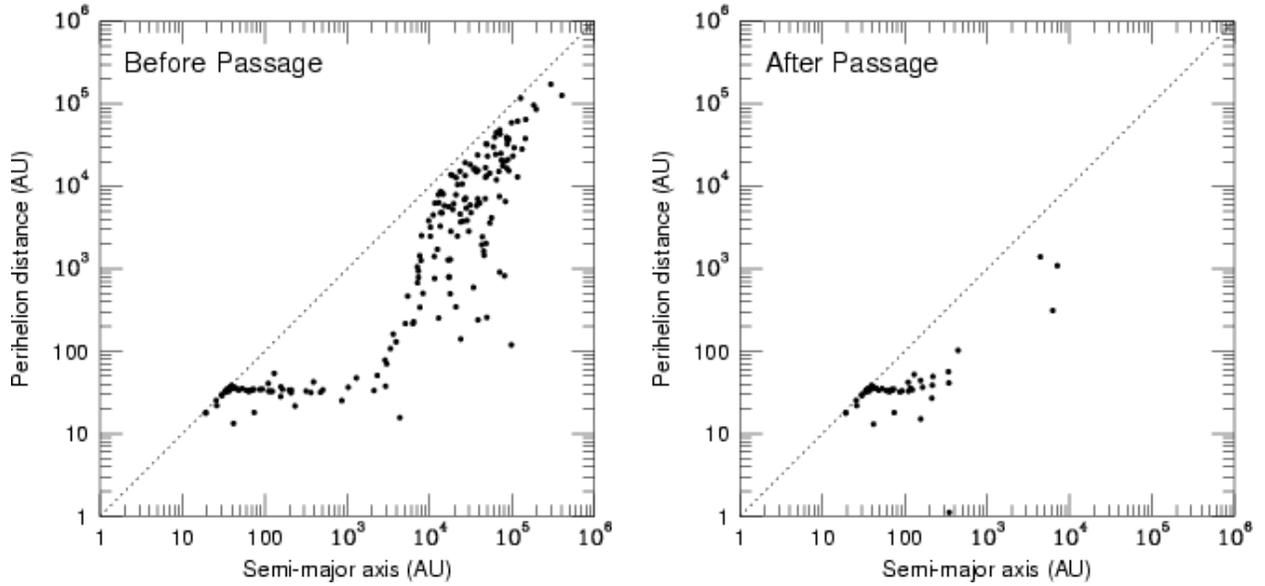}
\caption{The semi-major axis -- perihelion distance distribution of
  the Oort cloud before and after our nominal stellar passage with
  $q_\star\!=\!800\,$AU at $1\,$Gyr.  The left panel is taken directly
  from the simulations in Dones et al$.$~(2004).  The right panel
  shows the effect of such a passage.  Note that the Oort cloud is
  decimated. We conclude that the stellar encounter that emplaced
  2000~CR$_{105}$ and 2003~VB$_{12}$ on their current orbit occurred
  early in the history of the Solar System.}
\label{cr-starcr4_1e9} 
\end{figure}

This does not mean, however, that such a stellar encounter could have
occurred at any time during the history of the Solar System. This is
due to the fact that such a stellar encounter would have stripped any
Oort cloud population that existed at the time of the encounter.  This
is illustrated in Fig.~\ref{cr-starcr4_1e9}, which shows a before and
after snapshot of the scatter disk and Oort cloud which suffered our
nominal stellar passage with $q_\star\!=\!800\,$AU at $10^9\,$ years.
Note that practically all the objects with $a\!\gtrsim\!400\,$AU were
stripped from the system.  Also there is almost no material left in
the scattered disk to rebuild it.  Indeed, using methods developed in
LMD04, we find that in this case the Oort cloud would only contain 8\%
of the material that it would have if the encounter never happened.
That is, we can rule out an encounter this late because the Oort cloud
would be too anemic (see LMD04 for a more detailed discussion).  Using
the same techniques, we find that any encounter 100~Myr or earlier
does not effect the mass of the final Oort cloud, while an encounter
at 100~Myr results in an Oort cloud that is 50\% of the nominal one.

\section{Conclusions and discussion}

We have analyzed with numerical simulations four seemingly promising
mechanisms for explaining the origin of the peculiar extended
scattered disk object 2000~CR$_{105}$: (i) a high eccentricity phase
of Neptune, (ii) the existence of a rogue planet in the Kuiper belt or
in the scattered disk, (iii) the tide exerted by a massive and
dynamically cold trans-Neptunian disk, and (iv) the passage of a star
near the Solar System.  Of these, only the early passage of a
Solar-mass star at about 800~AU from the Sun appears satisfactory.
This is also the only scenario that we have studied that can easily
explain the origin of the newly found object 2003~VB$_{12}$.

Another scenario for the origin of extended scattered disk objects
has been proposed by Gomes (2003a, 2003b).  In it, a small fraction of
the objects in an early massive scattered disk population permanently
acquire a large perihelion distance during the outer migration of
Neptune.  This mechanism predicts that there should be large-$q$
objects with semi-major axis all over the range from $\sim\!50$ to
$\sim\!200\,$AU. 

The fact that the extended scattered disk bodies with the largest
perihelion distances, 2000~CR$_{105}$ and 2003~VB$_{12}$, both have
$a>200$~AU, argues against a scenario like that of Gomes.  Since
observational biases (given an object's perihelion distance and
absolute magnitude, and a survey's limiting magnitude of detection)
sharply favors the discovery of objects with small semi-major axis, we
believe that it would be unlikely that the first two discovered body
with $q>44$~AU had $a\!>\!200$~AU, if the real semi-major axis
distribution in the extended scattered disk were skewed toward small
$a$.  An yet, Gomes's mechanism and all of the mechanisms we have
studied, except for the passing star, lead to such a distribution.
Indeed, it is intuitive that creating a distribution where the objects
with the largest $a$ also have the largest $q$ requires a perturbation
`from the outside', whose magnitude decreases with decreasing
heliocentric distance.  Thus, there are not many alternatives to the
stellar encounter scenario.

In the current galactic environment, the closest stellar encounter
that should occur over the age of the Solar System is at $\sim 900$~AU
(Garcia-Sanchez et al., 2001). This should typically happen with a
star that is about 1/10th the mass of the Sun.  Therefore, the
encounter with a Solar-mass star at $\sim 800$~AU at early times
requires that the environment in which that Sun formed was
significantly denser, such as that of a stellar cluster (Bate et al.,
2003).  In the future the existence of 2000~CR$_{105}$, 2003~VB$_{12}$,
and their cohorts my supply important clues to the exact environment in
which the Sun and Solar system formed.

On a final note, since its discovery, passing stars have been used to
capture objects into the Oort cloud (Oort 1950).  Indeed, the process
invoked for the Oort cloud is identical to the one employed here.
Thus, if a passing star is indeed responsible for the formation of
objects like 2000~CR$_{105}$ and 2003~VB$_{12}$, then it is perhaps more
appropriate to characterize these objects as the inner edge of the
Oort cloud rather than the outer edge of the scattered disk.  Indeed,
recent simulations of the Oort cloud in a star cluster (Eggers~1997;
1998; Fernandez and Brunini~2000) actually produce objects like
2003~VB$_{12}$.

\acknowledgments We are very grateful to M. Brown for supplying
information on his new object, 2003~VB$_{12}$, a few days before the
official announcement.  HFL is grateful for funding from NASA's
Origins and PGG Programs.  We thank the CNRS-NSF exchange program for
supporting HFL's sabbatical at Observatoire de la C\^ote d'Azur, Nice,
France.

\section{References}

\begin{itemize}
\setlength{\itemindent}{-30pt}
\setlength{\labelwidth}{0pt}

\item[] Allen, R.~L., Bernstein, G.~M. \& Malhotra, R. 2002.
  Observational Limits on a Distant Cold Kuiper Belt. {\it Astron.
    J.}, {\bf 124}, 2949-2954.

\item[] Bate M.R., Bonnell I.A. and Bromm V. 2003. The formation of a 
star cluster: predicting the properties of stars and brown dwarfs.
{\it M.N.R.A.S.}, {\bf 339}, 577-599.

\item[] Brown M., et al$.$ 2004. {\it MPEC 2004-E45}.

\item[] Brown M. 2001.
The Inclination Distribution of the Kuiper Belt.
{\it Astron. J.}, {\bf 121}, 2804-2814.

\item[] Brunini A. and Melita M. (2002)
The existence of a planet beyond 50 AU
and the orbital distribution of the classical Edgeworth Kuiper 
belt objects. {\it Icarus}, {\bf 160}, 32-43.

\item[] Dones L. 2004. Simulations of the Formation of the Oort cloud
  I: The References Models. {\it Icarus}, submitted. 

\item[] Duncan, M. J., Levison, H. F. 1997,  scattered comet
disk and the origin of Jupiter family comets, Science, 276, 1670--1672.

\item[] Eggers, S., H.~U.~Keller, P.~Kroupa, and W.~J.~Markiewicz
  1997.\ Origin and dynamics of comets and star formation.\ {\it
    Planet. Space Sci.}  {\bf 45}, 1099--1104.
  
\item[] Eggers, S., H.~U.~Keller, W.~J.~Markiewicz, and P.~Kroupa
  1998.\ Cometary dynamics in a star cluster.\ {\it Astronomische
    Gesellschaft} meeting abstracts {\bf 14}, 5 (abstract).

\item[] Emel'yanenko V. V., Asher D. J., Bailey M. E. 2003.
A new class of trans-Neptunian objects in high-eccentricity orbits.
{\it MNRAS}, {\bf 338}, 443-451

\item[] Fernandez J. and Brunini A., 2000. The buildup of a tightly bound
  comet cloud around an early Sun immersed in a dense galactic environment:
  Numerical experiments. {\it Icarus}, {\bf 145}, 580--590.

\item[]Garcia-Sanchez J., Weissman P.R., Preston R.A.,
 Jones D.L.,  Lestrade J.F., Latham D.W.,
 Stefanik R.P., Paredes J.M. 2001. Stellar encounters with the solar system. {\it
    Astron. Astropys.}, {\bf 379}, 634-659. 

\item[] Gladman B., Holman M., Grav T., Kaavelars J.J., Nicholson P., Aksnes
  K., Petit J.M. 2001. Evidence for an extended scattered disk. {\it Icarus}, in press.

\item[] Gomes R.S. 2003a.  The origin of the Kuiper belt high inclination
  population. {\it Icarus}, {\bf 161}, 404-418. 

\item[] Gomes R.S. 2003b. A new model for the origin of the high-inclined
  TNOs. In press in {\it Proceedings of the First Decadal
Review of the Edgeworth-Kuiper Belt Meeting in Antofagasta}, Chile, to be
published in {\it Earth Moon and Planets}

\item[] Ida, S., Larwood, J. \& Burkert A. 2000.
Evidence for Early Stellar Encounters in the Orbital Distribution of Edgeworth-Kuiper Belt Objects.
ApJ., 528, 351--356.

\item[] Kobayashi H., Ida S. 2001. The Effects of a Stellar Encounter 
  on a Planetesimal Disk. {\it Icarus}, {\bf 153}, 416-429. 

\item[] Levison H.F., Duncan M.J. 1994, The long-term
dynamical behavior of short-period comets. {\it  Icarus}, 108:18--36.

\item[] Levison, H.F., \& Duncan, M. J. 1997, From the Kuiper Belt to Jupiter-Family Comets: The Spatial
               Distribution of Ecliptic Comets, Icarus, 127, 13--32.
\item[] Levison H.F., Morbidelli A. and Dones L. 2004.
               Forming the Outer Edge of the Kuiper Belt by a Stellar
               Encounter: Constraints from the Oort Cloud. {\it
                 Astron. J.}, submitted.

\item[] Melita M., Larwood J., Collander-Brown S, Fitzsimmons A., Williams
  I.P., Brunini A. 2002. The edge of the Edgeworth-Kuiper belt: stellar
  encounter, trans-Plutonian planet or outer limit of the primordial solar
  nebula? In {\it Asteroid, Comet, Meteors}, ESA Spec. Publ. series, in press.

\item[] Morbidelli, A., \& Valsecchi, G. B. 1997, Neptune
scattered planetesimals could have sculpted the primordial
Edgeworth--Kuiper belt, Icarus, 128, 464--468.

\item[] Morbidelli A. 2002 {\it Modern celestial mechanics: aspects of solar
    system dynamics}, Taylor and Francis, London.

\item[] Morbidelli A., Brown M. and Levison H. 2003. The Kuiper belt and its
  primordial sculpting. In press in {\it Proceedings of the First Decadal
Review of the Edgeworth-Kuiper Belt Meeting in Antofagasta}, Chile, to be
published in {\it Earth Moon and Planets}

\item[] Morbidelli A., Emel'yanenko V. and Levison H.F. 2004. Origin and
  orbital distribution of the trans Neptunian Scattered Disk. {\it MNRAS},
  submitted. 

\item[] Petit, J. M., Morbidelli, A., \& Valsecchi, G. B. 1999.
Large scattered planetesimals and the excitation of the small body
belts, Icarus, 141, 367-387.

\item[] Stern, S. A. 1996, On the Collisional Environment, Accretion Time
  Scales, and Architecture of the Massive, Primordial Kuiper
Belt., AJ, 112, 1203--1210.

\item[] Thomas F., Morbidelli A. 1996.
The Kozai resonance in the outer Solar System and the
dynamics of long--period comets,  {\it Celest. Mech.},{\bf 64}, 209--229. 

\item[] Thommes, E. W., Duncan, M. J. \& Levison, H. F. 1999.
The formation of Uranus and Neptune in the Jupiter-Saturn region of the Solar 
System. Nature, 402, 635--638.

\item[] Trujillo, C. A. and Brown, M.E. 2001.
The Radial Distribution of the Kuiper Belt.
AJ, 554, L95--L98.

\item[] Weissman, P.R., Levison H.F. 1997, The
population of the trans--neptunian region, in Pluto, D. J. Tholen
and S. A. Stern, Tucson: Univ. Ariz. Press. , 559--604

\end{itemize}

\clearpage

\clearpage

\clearpage

\clearpage

\clearpage

\end{document}